\documentclass[english]{article}
\usepackage{lmodern}

\usepackage[T1]{fontenc}
\usepackage[latin9]{inputenc}
\usepackage{geometry}
\usepackage{babel}
\usepackage{verbatim}
\usepackage{varioref}
\usepackage{amsmath}
\usepackage{amssymb}
\PassOptionsToPackage{normalem}{ulem}
\usepackage{ulem}
\usepackage[unicode=true,pdfusetitle,
 bookmarks=true,bookmarksnumbered=true,bookmarksopen=true,bookmarksopenlevel=1,
 breaklinks=true,pdfborder={0 0 0},backref=false,colorlinks=false]
 {hyperref}

\makeatletter
\newcommand{\code}[1]{\texttt{#1}}

\@ifundefined{date}{}{\date{}}
\makeatother

\begin{document}

\title{\noindent Induction, Coinduction, and Fixed Points\\
in Programming Languages (PL) Type Theory}

\author{Moez A. AbdelGawad\\
\code{moez@cs.rice.edu}}
\maketitle
\begin{abstract}
Recently we presented a concise survey of the formulation of the induction
and coinduction principles, and some concepts related to them, in
programming languages type theory and four other mathematical disciplines.
The presentation in type theory involved the separate formulation
of these concepts, first, in the theory of types of functional programming
languages and, next, in the theory of types of object-oriented programming
languages.

In this article we show that separating these two formulations helps
demonstrate some of the fundamental differences between structural
subtyping, predominant in functional programming languages, and nominal
subtyping, predominant in object-oriented programming languages---including
differences concerning type\emph{ }negation and concerning the existence
of inductive\emph{ }types, of coinductive\emph{ }types, and of approximations
thereof. In the article we also motivate mutual coinduction and mutual\emph{
}coinductive\emph{ }types, and their approximations, and we discuss
in brief the potential relevance of these concepts to object-oriented
programming (OOP) type theory.
\end{abstract}

\section{\label{sec:Introduction}Introduction}

Applications of the induction and coinduction principles, and of concepts
related to them, in scientific fields are plenty, including ones in
economics and econometrics~\cite{McLenn2018}, in mathematical physics~\cite{Penrose2004},
in computer science, and in many other areas of mathematics itself.
Their applications in computer science, in particular, include ones
in programming language semantics---which we touch upon in this article---as
well as in relational database theory (\emph{e.g.}, recursive or iterated
joins) and in concurrency theory.\footnote{For more details on the use of induction, coinduction, and fixed points
in computer science, and for concrete examples of how they are used,
the reader is invited to check relevant literature on the topic, \emph{e.g.},
\cite{Greiner92,TAPL,Bertot2004,Sangiorgi2012,Kozen2016}.}

Fixed points, induction and coinduction have been formulated and studied
in various subfields of mathematics, usually using different vocabulary
in each field. The interested reader is invited to check our concise
comparative survey in~\cite{AbdelGawad2018f}.

Of particular interest to programming languages researchers, in this
article we give attention to the fundamental conceptual difference
between \emph{structural} type theory, where type equality and type
inclusion are defined based on type structures but not on type names,
and \emph{nominal} type theory, where type equality and type inclusion
are defined based on the names of types in addition to their structures.
Given their difference in how each particularly defines the \emph{inclusion}
relation between types (\emph{i.e.}, the subtyping relation), the
conceptual difference between structural typing and nominal typing
expresses itself, prominently, when fixed points and related concepts
are formulated in each of structural type theory and nominal type
theory.

As such, in this article we present the formulation of these concepts
in the theory of types of functional programming languages (which
is largely structurally-typed) in~$\mathsection$\ref{sub:(Co)Inductive-Types}~(\nameref{sub:(Co)Inductive-Types}),
then we follow that by presenting their formulation in the type theory
of object-oriented programming languages (which is largely nominally-typed)
in~$\mathsection$\ref{sub:Object-Oriented-Type-Theory}~(\nameref{sub:Object-Oriented-Type-Theory}).

In~$\mathsection$%
7 of~\cite{AbdelGawad2018f} we summarized the article by presenting
tables that collect the formulations given in the article. Based on
the tabular comparison in~$\mathsection$%
7 of~\cite{AbdelGawad2018f} and based on the discussions in~$\mathsection$\ref{sec:Programming-Languages-Theory}
(and in~$\mathsection$%
3 of~\cite{AbdelGawad2018f}), we discuss in~$\mathsection$\ref{sec:Struct-vs-Nom}~(\nameref{sec:Struct-vs-Nom})
some of the fundamental differences between structural typing and
nominal typing, particularly ones related to type negation and coinductive
types, and we also discuss some consequences of these fundamental
differences.

\section{\label{sec:Programming-Languages-Theory}Programming Languages Theory}

Given the `types as sets' view of types in programming languages,
this section builds on the \emph{set}-theoretic presentation in~$\mathsection$%
3 of~\cite{AbdelGawad2018f} to present the induction and coinduction
principles using the jargon of programming languages type theory.
The presentation allows us to demonstrate and discuss the influence
structural and nominal typing have on the theory of type systems of
functional programming languages (which mostly use structural typing)
and object-oriented programming languages (which mostly use nominal
typing).

\subsection[FP Type Theory]{\label{sub:(Co)Inductive-Types}Inductive and Coinductive Functional
Data Types}

\subsubsection{Formulation}

Let $\mathbb{D}$ be the set of \emph{structural} types in functional
programming.\footnote{By construction/definition, the poset of structural types $\mathbb{D}$
under the inclusion/structural subtyping ordering relation is always
a complete lattice. This point is discussed in more detail below.}

\medskip{}

Let $\subseteq$ (`is a subset/subtype of') denote the \emph{structural
subtyping/inclusion} relation between structural data types, and let
$:$ (`has type/is a member of/has structural property') denote
the \emph{structural typing} relation between structural data values
and structural data types.

Now, if $F:\mathbb{D}\rightarrow\mathbb{D}$ is a polynomial (with
powers) datatype constructor%
\emph{}\footnote{That is, $F$ is one of the $+$, $\times$, or $\rightarrow$ data
type constructors (\emph{i.e.}, the summation/disjoint-union/variant
constructor, the product/record/labeled-product constructor, or the
continuous-function/exponential/power constructor, respectively) or
is a composition of these constructors. By their definitions in domain
theory~\cite{Scott82,PlotkinDomains83,GunterHandbook90,KahnConDoms93,Abramsky94,Gierz2003,DomTheoryIntro},
these structural datatype constructors, and their compositions, are
monotonic (also called \emph{covariant}) datatype constructors (except
for the first type argument of $\rightarrow$, for which $\rightarrow$
is an \emph{anti-monotonic}/\emph{contravarian}t constructor, but
that otherwise ``behaves nicely''~\cite{MPS}).}%
, \emph{i.e.}, if 
\[
\forall S,T\in\mathbb{D}.S\subseteq T\implies F\left(S\right)\subseteq F\left(T\right),
\]
then an inductively-defined type/set $\mu_{F}$, the \emph{smallest
$F$-closed} \emph{set}, exists in $\mathbb{D}$, and $\mu_{F}$ is
also the \emph{smallest fixed point} of $F$, and a coinductively-defined
type/set $\nu_{F}$, the \emph{largest $F$-consistent} \emph{set},
exists in $\mathbb{D}$, and $\nu_{F}$ is also the \emph{largest
fixed point} of $F$.\footnote{See Table~1 of~\cite{AbdelGawad2018f} for the definitions of $\mu_{F}$
and $\nu_{F}$.}%

Further, for any type $P\in\mathbb{D}$ (where $P$, as a structural
type, expresses a structural property of data values) we have:
\begin{itemize}
\item (\emph{structural induction, and recursion})\phantom{co} $F\left(P\right)\subseteq P\implies\mu_{F}\subseteq P$\medskip{}
\\
(\emph{i.e., }$\forall p:F\left(P\right).p:P\implies\forall p:\mu_{F}.p:P$),\medskip{}
\\
which, in words, means that if the (structural) property $P$ is \emph{preserved}
by $F$ (\emph{i.e.}, if $P$ is $F$-closed), then all data values
of the inductive type $\mu_{F}$ have property $P$ (\emph{i.e.},
$\mu_{F}\subseteq P$).\\
Furthermore, borrowing terminology from category theory (see~$\mathsection$%
6 of~\cite{AbdelGawad2018f}), a \emph{recursive} function $f:\mu_{F}\rightarrow P$
that maps data values of the inductive type $\mu_{F}$ to data values
of type $P$ (\emph{i.e.}, having structural property $P$) is the
unique \emph{catamorphism} (also called a \emph{fold}) from $\mu_{F}$
to $P$ (where $\mu_{F}$ is viewed as an initial $F$-algebra and
$P$ as an $F$-algebra), and
\item (\emph{structural coinduction, and corecursion}) $P\subseteq F\left(P\right)\implies P\subseteq\nu_{F}$\medskip{}
\emph{}\\
(\emph{i.e., }$\forall p:P.p:F\left(P\right)\implies\forall p:P.p:\nu_{F}),$\medskip{}
\\
which, in words, means that if the (structural) property $P$ is \emph{reflected}
by $F$ (\emph{i.e.}, if $P$ is $F$-consistent), then all data values
that have property $P$ are data values of the coinductive type $\nu_{F}$
(\emph{i.e.}, $P\subseteq\nu_{F}$).\\
Furthermore, borrowing terminology from category theory, a \emph{corecursive}
function $f:P\rightarrow\nu_{F}$ that maps data values of type $P$
(\emph{i.e.}, having structural property $P$) to data values of the
coinductive type $\nu_{F}$ is the unique \emph{anamorphism} from
$P$ to $\nu_{F}$ (where $P$ is viewed as an $F$-coalgebra and
$\nu_{F}$ as a final $F$-coalgebra).
\end{itemize}

\subsubsection{Notes}
\begin{itemize}
\item To guarantee the existence of $\mu_{F}$ and $\nu_{F}$ in $\mathbb{D}$
for all type constructors $F$, and hence to guarantee the ability
to reason easily---\emph{i.e.}, inductively and coinductively---about
functional programs, the domain $\mathbb{D}$ of %
types in functional programming is \emph{deliberately} constructed
to be a complete lattice under the inclusion ordering%
. This is achieved by limiting the type constructors used in constructing
$\mathbb{D}$ and over $\mathbb{D}$ to \emph{structural} type constructors
only (\emph{i.e.}, to the constructors $+$, $\times$, $\rightarrow$
and their compositions, in addition to basic types such as \code{Unit},
\code{Bool}, \code{Top}, \code{Nat} and \code{Int}).

\begin{itemize}
\item For example, the \emph{inductive} type of lists of integers in functional
programming is defined structurally (\emph{i.e.}, using $+$, $\times$,
and structural induction) as 
\[
L_{i}\simeq\mathtt{Unit}+\mathtt{Int}\times L_{i},
\]
which defines the type $L_{i}$ as (isomorphic/equivalent to) the
summation of type \code{Unit} (which provides the value \code{unit}
as an encoding for the empty list) to the product of type \code{Int}
with type $L_{i}$ itself.
\item In fact the three basic types \code{Bool}, \code{Nat} and \code{Int}
can also be defined structurally. For example, in a functional program
we may structurally define type \code{Bool} using the definition
$\mathtt{Bool}\simeq\mbox{\ensuremath{\mathtt{Unit}}+\ensuremath{\mathtt{Unit}}}$
(for \code{false} and \code{true}), structurally define type \code{Nat}
using the definition $\mathtt{Nat}\simeq\mbox{\ensuremath{\mathtt{Unit}}+\ensuremath{\mathtt{Nat}}}$
(for 0 and the successor of a natural number), and, out of other equally-valid
choices, structurally define type \code{Int} using the definition
$\mathtt{Int}\simeq\mbox{\ensuremath{\mathtt{Nat}}+\ensuremath{\mathtt{Unit}}+\ensuremath{\mathtt{Nat}}}$
(for negative integers, zero, and positive integers).%

\end{itemize}
\end{itemize}

\subsubsection{References}

See~\cite{TAPL,Gapeyev2002,Bertot2004,Geldrop2009}.

\subsection[OOP Type Theory]{\label{sub:Object-Oriented-Type-Theory}Object-Oriented Type Theory}

The accurate and precise understanding of the generic\emph{ }subtyping
relation in mainstream OOP languages such as Java, C\#, C++, Kotlin
and Scala, and the proper mathematical modeling of the OO subtyping
relation in these languages, is one of our main research interests.
Due to the existence of features such as wildcard types, type erasure,
and bounded generic classes (where classes\footnote{The notion of \emph{class} in this article includes that of an \code{abstract class},
of an \code{interface}, and of an \code{enum} in Java~\cite{JLS18}.
It also includes similar ``type-constructing'' constructs in other
nominally-typed OO languages, such as \code{trait}s in Scala~\cite{Odersky14}.
And a \emph{generic} \emph{class} is a class that takes a type parameter
(An example is the generic interface \code{List} in Java---that models
lists/sequences of items---whose type parameter specifies the type
of items in a list).} play the role of type constructors), the mathematical modeling of
the generic subtyping relation in mainstream OOP languages is a hard
problem that, in spite of much effort, seems to still have not been
resolved, at least not completely nor satisfactorily, up to the present
moment~\cite{Tate2011,Greenman2014,AbdelGawad2017a,AbdelGawad2017b,AbdelGawad2018b,AbdelGawad2018e}.

The majority of mainstream OO programming languages are class-based,
and subtyping ($<:$) is a fundamental relation in OO software development.
In industrial-strength OOP, \emph{i.e.}, in statically-typed class-based
OO programming languages such as Java, C\#, C++, Kotlin and Scala,
class names are used as type names, since class names---which objects
carry at runtime---are assumed to be associated with \emph{behavioral}
\emph{class contracts} by developers of OO software. Hence, the decision
of equality between types in these languages takes type names in consideration---hence,
nominal typing. In agreement with the nominality of typing in these
OO languages, the fundamental subtyping relation in these languages
is \emph{also} a nominal relation. Accordingly, subtyping decisions
in the type systems of these OO languages make use of the inherently-nominal
inheritance declarations (\emph{i.e., }that are \emph{explicitly}
declared between class \emph{names}) in programs written using these
languages.\footnote{Type/contract inheritance that we discuss in this article is the same
thing as the inheritance of behavioral interfaces (APIs) from superclasses
to their subclasses that mainstream OO software developers are familiar
with. As is empirically familiar to OO developers, the subtyping relation
in class-based OO programming languages is in one-to-one correspondence
with API (and, thus, type/contract) inheritance from superclasses
to their subclasses~\cite{Tempero:2013:PIJ:2524984.2525016}. Formally,
this correspondence is due to the nominality of the subtyping relation~\cite{NOOPsumm}.} (For a more detailed overview of nominal typing versus structural
typing in OOP see~\cite{OOPOverview13}.)

\subsubsection{Formulation}

Let $<:$ (`is a subtype of') denote the \emph{nominal subtyping}
relation between nominal data types (\emph{i.e.}, class types), and
let $:$ (`has type') denote the \emph{nominal typing} relation
between nominal data values (\emph{i.e.}, objects) and nominal data
types.

Further, let $\mathbb{T}$ be the set of nominal types in object-oriented
programming, ordered by the nominal subtyping relation, and let $F:\mathbb{T}\rightarrow\mathbb{T}$
be a type constructor over $\mathbb{T}$ (\emph{e.g.}, a generic class).\footnote{Unlike poset $\mathbb{D}$ in $\mathsection$\ref{sub:(Co)Inductive-Types}
(of structural types under the structural subtyping relation), poset
$\mathbb{T}$ (of nominal types under the nominal subtyping relation)
is \emph{not} guaranteed to be a complete lattice.}

A type $P\in\mathbb{T}$ is called an `\emph{$F$-supertype}' if
\uline{its $F$-image is a subtype of it}, \emph{i.e.}, if 
\[
F\left(P\right)<:P,
\]
and $P$ is said to be \emph{preserved} by $F$. (An $F$-supertype
is sometimes also called an \emph{$F$-closed type}, \emph{$F$-lower
bounded type}, \emph{$F$-large type}, \emph{inductive type}, or \emph{algebraic
type}). The \emph{root }or \emph{top} of the subtyping hierarchy,
if it exists (in $\mathbb{T}$), is usually called \code{Object}
or \code{All}, and it is an $F$-supertype for all generic classes
$F$. In fact the top type, when it exists, is the \emph{greatest
$F$-supertype} for all $F$.

A type $P\in\mathbb{T}$ is called an `\emph{$F$-subtype}' if \uline{it
is a subtype of its $F$-image}, \emph{i.e.}, if 
\[
P<:F\left(P\right),
\]
and $P$ is said to be \emph{reflected} by $F$. (An $F$-subtype
is sometimes also called an \emph{$F$-consistent type}, \emph{$F$-(upper)
bounded type}\footnote{From which comes the name \emph{$F$-bounded generics} in object-oriented
programming.}, \emph{$F$-small type}, \emph{coinductive type}, or \emph{coalgebraic
type}). The \emph{bottom} of the subtyping hierarchy, if it exists
(in $\mathbb{T}$), is usually called \code{Null} or \code{Nothing},
and it is an $F$-subtype for all generic classes $F$. In fact the
bottom type, when it exists, is the \emph{least $F$-subtype} for
all $F$.

A type $P\in\mathbb{T}$ is called a \emph{fixed point} (or `\emph{fixed
type}') of $F$ if \uline{it is equal to its $F$-image}, \emph{i.e.},
if 
\[
P=F\left(P\right).
\]
As such, a fixed point of $F$ is simultaneously an $F$-supertype
and an $F$-subtype. (Such fixed types/points are rare in OOP practice).

\medskip{}

Now, if $F$ is\emph{ }a \emph{covariant} generic class (\emph{i.e.},
a types-generator)\footnote{\label{fn:Gen-Cov-Int}Generic classes in Java are in fact \emph{always}
monotonic/covariant, not over types ordered by subtyping but over
\emph{interval types }ordered by \emph{containment}. (See~\cite{AbdelGawad2018c}.)
In particular, for \emph{any} generic class $F$ in Java we have 
\[
I_{1}\sqsubseteq I_{2}\implies F\left(I_{1}\right)<:F\left(I_{2}\right),
\]
meaning that if interval type $I_{1}$ is a \emph{subinterval-of}
(or \emph{contained-in}) interval type $I_{2}$ then the instantiation
of generic class $F$ with $I_{1}$ (\emph{i.e.}, the parameterized
type $F\left(I_{1}\right)$, usually written as $F\left\langle I_{1}\right\rangle $)
is always a \emph{subtype-of} the instantiation of class $F$ with
$I_{2}$ (\emph{i.e.}, of the parameterized type $F\left(I_{2}\right)$,
usually written as $F\left\langle I_{2}\right\rangle $). As such,
a generic class in Java is \emph{not }exactly an endofunction over
types but is rather what may be called an ``indirect endofunction,''
since it generates/constructs types not directly from types but from
interval types that are themselves \emph{derived} \emph{from} types
(see~\cite{AbdelGawad2018b,AbdelGawad2018c}), and the generic class,
as a function, is monotonic/covariant with respect to the containment
relation over these interval types (\emph{i.e.}, it ``generates a
subtype when provided with a subinterval'').

}%
, \emph{i.e.}, if 
\[
\forall S,T\in\mathbb{T}.S<:T\implies F\left(S\right)<:F\left(T\right),
\]
and if $\mu_{F}$, the `\emph{least $F$-supertype}' exists in $\mathbb{T}$,
and $\mu_{F}$ is also the \emph{least fixed point} of $F$, and if
$\nu_{F}$, the `\emph{greatest $F$-subtype}', exists in $\mathbb{T}$,
and $\nu_{F}$ is also the \emph{greatest fixed point} of $F$,\footnote{See Table~2 of~\cite{AbdelGawad2018f} for the definitions of $\mu_{F}$
and $\nu_{F}$ in the (rare) case when $\mathbb{T}$ happens to be
a complete lattice.} then, for any type $P\in\mathbb{T}$ we have:
\begin{itemize}
\item (\emph{induction})\phantom{co} $F\left(P\right)<:P\implies\mu_{F}<:P$\medskip{}
\\
(\emph{i.e., }$\forall p:F\left(P\right).p:P\implies\forall p:\mu_{F}.p:P$),\medskip{}
\\
which, in words, means that if the contract (\emph{i.e.}, behavioral
type) $P$ is \emph{preserved} by $F$ (\emph{i.e.}, $P$ is an $F$-supertype),
then the inductive type $\mu_{F}$ is a subtype of $P$, and
\item (\emph{coinduction}) $P<:F\left(P\right)\implies P<:\nu_{F}$\medskip{}
\emph{}\\
(\emph{i.e., }$\forall p:P.p:F\left(P\right)\implies\forall p:P.p:\nu_{F}$),\medskip{}
\\
which, in words, means that if the contract (\emph{i.e.}, behavioral
type) $P$ is \emph{reflected} by $F$ (\emph{i.e.}, $P$ is an $F$-subtype),
then $P$ is a subtype of the coinductive type $\nu_{F}$.
\end{itemize}

\subsubsection{Notes}
\begin{itemize}
\item As discussed earlier and in $\mathsection\ref{sub:(Co)Inductive-Types}$,
in structural type theory type expressions express only structural
properties of data values, \emph{i.e., }how the data values of the
type are structured and constructed. In nominal type theory type names
are associated with formal or informal contracts, called \emph{behavioral}
\emph{contracts}, which express behavioral properties of the data
values (\emph{e.g.}, objects) in addition to their structural properties.

\begin{itemize}
\item To demonstrate, in a pure structural type system a record type that
has, say, one member (\emph{e.g.}, type \code{plane} $=\{$ \code{fly()}
$\}$, type \code{bird} $=\{$ \code{fly()} $\}$ and type \code{insect}
$=\{$ \code{fly()} $\}$) is semantically \emph{equivalent} to any
other type that has the same member (\emph{i.e.}, type \code{plane}
is equivalent to type \code{bird} and to type \code{insect})---in
other words, in a pure structural type system these types are `interchangeable
for all purposes'.
\item On the other hand, in a pure \emph{nominal} type system any types
that have the same structure but have different names (\emph{e.g.},
types \code{plane}, \code{bird} and \code{insect}) are considered
\emph{distinct} types that are \emph{not} semantically equivalent,
since their different names (\emph{e.g.}, `\code{plane}' versus
`\code{bird}' versus `\code{insect}') imply the possibility,
even likelihood, that data values of each type maintain \emph{different}
behavioral contracts, and thus of the likelihood of \emph{different}
use considerations for the types and their data values.\footnote{For another example, a \code{float} used for monetary values (\emph{e.g.},
in finanicial transactions) should normally \emph{not} be confused
with (\emph{i.e.}, equated to) a \code{float} used for measuring
distances (\emph{e.g.}, in scientific applications). Declaring

\begin{center}
\code{\textbf{type} money=float}\\
\code{\textbf{~~~type} distance=float}
\par\end{center}

\noindent does not help in a purely structural type system, however,
since the types \code{float}, \code{money}, and \code{distance}
are structurally equivalent. On the other hand, in a purely nominal
type system the declarations of types \code{money} and \code{distance}
\emph{do} have the desired effect, since the non-equivalence of the
types is implied by their different \emph{names}.}\textsuperscript{,}\footnote{Further, when (1) the functional components of data values are (\emph{mutually})
\emph{recursive}, which is typical for methods of objects in OOP~\cite{AbdelGawad2017c},
and when (2) data values (\emph{i.e.}, objects) are \emph{autognostic}
data values\emph{ }(\emph{i.e.}, have a notion of \code{self/this},
which is an essential feature of mainstream OOP~\cite{cook-revisited})---which
are two features of OOP that necessitate \emph{recursive types}---then
the semantic differences between nominal typing and structural typing
become even more prominent, since type names and their associated
contracts gain more relevance as expressions of the richer recursive
behavior of the more complex data values. (For more details, see~\cite{AbdelGawad2015}
and \cite[$\mathsection$19.3]{TAPL}.)}
\end{itemize}
\item In industrial-strength OO programming languages (such as Java, C\#,
C++, Kotlin and Scala) where types are nominal types rather than structural
ones and, accordingly, where subtyping is a nominal relation, rarely
is poset $\mathbb{T}$ a lattice under the subtyping relation $<:$,
let alone a complete lattice. Further, \emph{many} type constructors
(\emph{i.e.}, generic classes) in these languages are not covariant.
As such, $\mu_{F}$ and $\nu_{F}$ rarely exist in $\mathbb{T}$.\footnote{Annihilating the possibility of reasoning inductively or coinductively
about nominal OO types.}%
{} Still, the notion of a pre-fixed point (or of an $F$-algebra) of
a generic class $F$ and the notion of a post-fixed point (or of an
$F$-coalgebra) of $F$, under the names $F$-supertype and $F$-subtype
respectively, do have relevance in OO type theory%
, \emph{e.g.}, when discussing $F$-bounded generics~\cite{AbdelGawad2018e}.\footnote{\label{fn:gfsub-lfsup}In fact, owing to our research interests (hinted
at in the beginning of~$\mathsection$\ref{sub:Object-Oriented-Type-Theory}),
inquiries in~\cite{AbdelGawad2018e} have been a main initial motivation
for writing this note/article. In particular, we have noted that if
$F$ is a generic class in a Java program then a role similar to the
role played by the coinductive type $\nu_{F}$ is played by the wildcard
type \code{F<?>}, since, by the subtyping rules of Java (discussed
in~\cite{AbdelGawad2018e}, and illustrated vividly in earlier publications
such as~\cite{AbdelGawad2018b,AbdelGawad2017a}), every $F$-subtype
(\emph{i.e.}, every parameterized type constructed using $F$---called
an \emph{instantiation} of $F$---and every subtype thereof) \emph{is}
a subtype of the type \code{F<?>}. On the other hand, in Java there
is not a non-\code{Null} type (not even type \code{F<Null>}; see~\cite{AbdelGawad2018e})
that plays a role similar to the role played above by the inductive
type $\mu_{F}$ (\emph{i.e.}, a type that is a subtype of all $F$-supertypes,
which are all instantiations of $F$ and all supertypes thereof).
This means that in Java greatest \emph{post-}fixed points (\emph{i.e.},
greatest $F$-subtypes) that are \emph{not} greatest fixed points
do exist, while non-bottom least \emph{pre-}fixed points (\emph{i.e.},
least $F$-supertypes) do not exist. Also, since $\mathbb{T}$ is
rarely a complete lattice, greatest fixed points, generally-speaking,
do not exist in Java, neither do least fixed points. These same observations
apply more-or-less to other nominally-typed OOP languages similar
to Java, such as C\#, C++, Kotlin and Scala. (See further discussion
in Footnote~18 in~$\mathsection$ of~\cite{AbdelGawad2018f}.)}
\item While not immediately obvious (nor widely-known), but, as a \emph{set}
of pairs of types, the subtyping relation in nominally-typed OOP
is in fact a \emph{coinductive} set, \emph{i.e.}, is a coinductively-defined
subset of the set of all pairs of class types~\cite[Ch. 21]{TAPL}.
That is because subtyping between two types holds in nominally-typed
OOP as long as \emph{there is no }(\emph{finite})\emph{ reason for
it not} to hold. The following is in fact how \code{javac}, the standard
Java compiler, type-checks Java programs: During type checking \code{javac}
assumes that a subtyping relation between two given types holds unless
the type checker can (finitely) prove, using the explicitly specified
subtyping declarations in a Java program, that the relation cannot
hold.\footnote{The subtyping relation in Java seems to be even a little bit more
complex than a coinductive set. Due to the existence of wildcard/interval
types, the Java subtyping relation, together with the containment
relation between wildcard/interval types~~\cite{AbdelGawad2018c},
seems to be an instance of a \emph{mutually coinductive} \emph{set}
(\emph{i.e.}, the coinductive counterpart of a \emph{mutually} \emph{inductive}
\emph{set}, which we did not get to discuss in this article but may
do in the future.)}

\begin{itemize}
\item In particular, to the best of our knowledge, the Java language specification
does \emph{not }stipulate that the subtyping relation holds between
two types \emph{only} if the relation can be (finitely) proven to
hold between the two types (see~\cite[$\mathsection$4.10]{JLS18}).
This missing ``disclaimer'', which is usually stipulated in the
definition of similar relations but that seems to be intentionally
missing in the definition of the subtyping relation between class
types (\emph{a.k.a.}, reference types) in Java, allows for the subtyping
relation in Java to be coinductive.
\end{itemize}
\end{itemize}

\subsubsection{References}

See~\cite{CanningFbounded89,BruceBinary94,Brandt1998,Baldan1999,Gapeyev2002,KennedyDecNomVar07,Tate2011,AbdelGawad2018e}.

\section[Struct. Type Theory vs. Nom. Type Theory]{\label{sec:Struct-vs-Nom}Structural Type Theory versus Nominal Type
Theory}

\subsection{Existence of Fixed Points}

The discussion in $\mathsection$\ref{sub:(Co)Inductive-Types}, together
with that in $\mathsection\mathsection$%
3 and~%
5 of~\cite{AbdelGawad2018f}, demonstrates that FP type theory, with
its structural types and structural subtyping rules being motivated
by mathematical reasoning about programs (using induction or coinduction),
is closer in its flavor to set theory (and first-order logic/predicate
calculus), since structural type theory assumes and \emph{requires}
the existence of fixed points $\mu_{F}$ and $\nu_{F}$ in $\mathbb{D}$
for all type constructors $F$. (For a discussion of  the importance
of structural typing in FP see~\cite{Hughes89,MacQueenMLOO02} and~\cite[$\mathsection$19.3]{TAPL}.)

On the other hand, the discussion in $\mathsection$\ref{sub:Object-Oriented-Type-Theory},
together with that in $\mathsection\mathsection$%
6 and~%
2 of~\cite{AbdelGawad2018f}, demonstrates that OOP type theory,
with its nominal types and nominal subtyping being motivated by the
association of nominal types with behavioral contracts, is closer
in its flavor to category theory and order theory, since nominal type
theory does \emph{not} assume or require the existence of fixed points
$\mu_{F}$ and $\nu_{F}$ in $\mathbb{O}$ for all type constructors
$F$. (%
For a discussion of why nominal typing and nominal subtyping matter
in OOP see~\cite{AbdelGawad2015} and~\cite[$\mathsection$19.3]{TAPL}.)

As such, we conclude that%
{} the theory of data types of functional programming languages%
{} is more similar in its views and its flavor to the views and flavor
of set theory and first-order logic, while the theory of data types
of object-oriented programming languages%
{} is more similar in its views and its flavor to those of category
theory and order theory. This conclusion adds further supporting evidence
to our speculation (\emph{e.g.}, in~\cite{AbdelGawad2017a,AbdelGawad2017b})
that category theory is more suited than set theory for the accurate
understanding of mainstream object-oriented type systems.

\subsection{\label{sub:OO-Type-Negation}Type Negation, Coinductive Nominal Types,
and Free Types}

In $\mathsection$%
3 of~\cite{AbdelGawad2019a} we noted that negation of a mathematical
object (\emph{e.g.}, a logical statement, a set, or a type) is useful
in defining coinductive objects. We also noted that negation in set
theory amounts to set complementation, which is useful in defining
coinductive sets in terms of inductive sets. We wondered also about
defining negation in other categories, such as the category of structural
types (ordered by structural subtyping) in functional programming
languages and the category of nominal types (ordered by nominal subtyping)
in object-oriented programming languages. We resume this discussion
here.

In object-oriented programming for example (see~$\mathsection$\ref{sub:Object-Oriented-Type-Theory}),
we may try to define negation of a class type as follows. First, let's
define an ``implication'' type (which may also be called an `exponential
type') from type $x$ to type $y$ as 
\begin{equation}
\left(x\Rightarrow y\right)\doteq\underset{x\wedge a<:y}{\bigvee a}\label{eq:imp-ord}
\end{equation}
\emph{i.e.}, as the join of all types whose meet with type $x$ is
a subtype of type $y$,\footnote{Formula~(\ref{eq:imp-ord}) comes from the study of Heyting algebras
(which model constructive/intuitionistic logic). The formula is valid
for defining complementation in set theory (where join $\vee$ is
interpreted as set union $\cup$, meet $\wedge$ is interpreted as
set intersection $\cap$, $\leq$ is interpreted as set inclusion
$\subseteq$, and $\bot$ is interpreted as the empty set $\phi$),
since the inclusion lattice in set theory is a Boolean algebra and
every Boolean algebra is a Heyting algebra.

As such, in set theory the complement $\lnot x$ of a set $x$ (as
a \emph{negation} of $x$) can also be defined as
\[
\lnot x\doteq\left(x\Rightarrow\phi\right)\doteq\underset{x\cap a=\phi}{\bigcup a}
\]
\emph{i.e.}, as the union of all subsets of $U$ that are disjoint
with respect to $x$. This definition, although it does not (explicitly)
mention $U$, is equivalent to the standard definition of set complementation,
namely the definition 
\[
\lnot x\doteq U-x\doteq U\backslash x.
\]
} then (as done in functional programming) let's define negation of
a type $x$ as 
\[
\lnot x\doteq\left(x\Rightarrow\bot\right)
\]
which in other words means defining type negation as 
\[
\lnot x\doteq\underset{x\wedge a=\bot}{\bigvee a},
\]
\emph{i.e.}, defining the negation of a type as \emph{the join of
all types parallel to the type},\footnote{\label{fn:parallel}The term \emph{parallel} here is used in an order-theoretic
sense. In an ordered set $\left(P,\leq\right)$ two elements $a,b\in P$
are parallel if neither $a\leq b$ holds nor $b\leq a$ holds (\emph{i.e.},
$a$ and $b$ are ``independent elements'' of $P$). This is usually
denoted by writing $a\parallel b$~\cite{Davey2002,OrderedSets05,Rom2008}.
(Although not widely-known, but \LaTeX{} has a command \code{\textbackslash{}parallel}
for inputting the symbol $\parallel$).} or more precisely as the join (``lub'') of those types whose meet
with the type is the bottom type.

As such, noting (see~$\mathsection$\ref{sub:Object-Oriented-Type-Theory})
that $\bot$ actually denotes type \code{Null} (sometimes also called
\code{Nothing} or \code{Void}) and $\top$ denotes type \code{Object}
(or \code{All}) then the negation of $\top$ is $\bot$ (since $\bot$
is the \emph{only} type whose meet with $\top$ is equal to $\bot$),
and the negation of $\bot$ is $\top$ (since the meet of $\bot$
and \emph{every} type, including $\top$, is $\bot$).

However, due to the general lack of (true) union types in most OOP
languages\footnote{In fact the \code{lub()} function---defined in the Java language
specification---does not compute the least upper bound of a set of
types (which itself is an approximation of the true union of the set
of types) but it `only \emph{approximates} a least upper bound'~\cite[$\mathsection$4.10.4]{JLS18}.}, for most types (\emph{i.e.}, ones other than $\top$ and $\bot$)
the negation of a nominal type $x$ will usually be type $\top$.
This makes the negation of class types, if defined as suggested above,
not quite interesting or useful. (Also, as observed in~$\mathsection$\ref{sub:Object-Oriented-Type-Theory},
the subtyping relation in mainstream OOP languages rarely has fixed
points of type generators/constructors).

An alternative way to defining the negation of a type in OOP in a
more genuine object-oriented way (\emph{i.e.}, not via defining implication/exponential
types) is to allow OO developers to choose which types they wish to
negate, rather than trying to define a negation ``automatically''
for every type. An OO developer may define a ``negation'' of a type
(in a genuine OO way, if they wished to have such a negative type)
by them simply extending the negative type from a supertype of the
negated type (or from one of its supertypes, depending on which part
of the inherited behavioral contract is being negated).

For example, say we have class \code{Window} that is extended by
class \code{ColoredWindow}. An OO developer can define the negation
of \code{ColoredWindow} by declaring a class \code{NonColoredWindow}
that extends class \code{Window}. (As is standard in nominal typing,
it is up to the developer to ensure that a \code{NonColoredWindow}
is indeed not colored, and in fact also that a \code{ColoredWindow}
is indeed one.)

Negative types, according to this genuine nominal/OO way of defining
them, are totally under the control of software developers, unlike
the case in structural/FP typing. It may be noted also that type negation
in a genuine OO way is not exclusive. For example, generally-speaking
there is nothing that prevents the OO developer from declaring yet
a third subclass of class \code{Window} that is not a \code{ColoredWindow}
yet is also not a \code{NonColoredWindow}. (That third class may
or may not be useful. The option of declaring it, or not, is available
to the developer however. In other words, it is the developer, not
the language, who makes the decision as to defining the class or not,
based on his or her need for the class.)

The more genuinely OO way of defining negative types presented above,
arguably, is better than the FP/structural way (via $T\Rightarrow\bot$),
since it gives more control and offers more flexibility to developers.
It should be noted, though, that there is a similarity between both
ways: they both define type negation by depending (directly, in OOP,
and indirectly, in FP) on `parallel types' (see Footnote~\vref{fn:parallel}).

Having considered negative nominal types, now what about \emph{coinductive
nominal types}? Understanding these types (as counterparts of coinductive
structural types in structural type theory and of coinductive sets
in set theory) was the initial motivation for us considering type
negation in the first place (see $\mathsection$%
3 of~\cite{AbdelGawad2019a}). The starting point for considering
such (strange-named, but familiar we assert) types will be Equation~%
(1) of~\cite{AbdelGawad2019a}, which, let's recall, is the equation
that defines coinductive sets (in terms of inductive ones and negation)
as
\[
\nu_{F}=\lnot\mu_{\lnot F\lnot}.
\]

Equation~%
(1) uses three negations, and we have just defined an intuitive notion
of type negation for nominal types. Our view of coinductive nominal
types will be a combination of both. In particular, we take each negation
in Equation~%
(1) \emph{separately} to define a negative nominal type (that may
or may not exist). As such, for a particular generic class $F$ (\emph{i.e.},
$F$ is a nominal type generator) we define its corresponding coinductive
nominal type $\nu_{F}$ (if it exists) as the negative of (an approximation
of, since nominal types do \emph{not} have fixed points) the inductive
type $\mu_{F^{\delta}}$ of a ``dual'' generic class $F^{\delta}$
(which, also, may or may not exist) that is applied to negations of
types (if all these negations exist) that can be passed to the generic
class $F$ as type arguments (!!).

Sounds too complex? (Due to the three negations?) Rest assured, in
conclusion, and to make a long story short, intuitively-speaking the
sought after \emph{coinductive nominal type} $\nu_{F}$ is roughly
(\emph{i.e.}, is approximated by) the familiar type $F\left\langle \mathsf{?}\right\rangle $---an
observation we earlier made (in Footnote~\vref{fn:gfsub-lfsup})
without any discussion of type negation. In other words, we have
\[
\nu_{F}\approx F\left\langle \mathsf{?}\right\rangle 
\]
 (again due to the absence of fixed points of generic classes in nominal
type theory, a true coinductive nominal type in fact never exists).
The type $F\left\langle \mathsf{?}\right\rangle $ is also called
the `\emph{free type}' corresponding to the generic class $F$.\footnote{The name `free type' comes from category theory. The concept of
a free type corresponding to a generic class is similar (in a precise
category theoretic sense) to the \emph{free monoid} corresponding
to a set and to the \emph{free category} (a \emph{quiver}) corresponding
to a graph~\cite{spivak2014category}. The free type corresponding
to a generic class is defined as the parameterized type formed by
instantiating the generic class with the wildcard type \code{?}.
For example, the free type corresponding to generic class \code{List}
is the type \code{List<?>}. Free types and (Java) \emph{type} \emph{erasure}
both form an adjunction---they are a pair of adjoint functors between
`classes \& subclassing' and `types \& subtyping'. (In order theory,
an adjunction is also called a Galois connection. See~\cite{AbdelGawad2017b}
for more details on this adjunction.) Based on the discussion in the
main text, free types and coinductive nominal types agree, \emph{i.e.},
are the same concept. More precisely, free types are the best approximation
of (true) coinductive nominal types. In category-theoretic terms free
types are, precisely, instances of final coalgebras. (See Table~2
of~\cite{AbdelGawad2018f}.)}

In light of the discussion (in $\mathsection$%
3 of~\cite{AbdelGawad2019a}) of the intuitions behind coinductive
sets (``good/con-structible/con\-sistent vs. bad/in\-con\-structible/in\-con\-sistent'',
and finitely-constructible vs. infinitely-constructible) and that
coinductive sets contain \emph{all} elements (all data values, in
case of coinductive structural data types) that can be constructed
using a generator (\emph{i.e.}, ones that do not ``break the rules''
of the gene\-rator/con\-structor), it is not surprising that, for
a generic class $F$, the free type $F\left\langle \mathsf{?}\right\rangle $
is the best approximation of the coinductive nominal type $\nu_{F}$.
That is because the free type $F\left\langle \mathsf{?}\right\rangle $
is intuitively understood (\emph{e.g.}, by OO software developers
and OO language designers) as the type that contains \emph{all} objects
that can be instances of (recall Java's \code{instanceof} operator)
the generic class $F$ (\emph{i.e.}, contains all objects that can
be constructed using the generic class $F$). Further, in light of
OOP \emph{not} having a goal of inductive logical reasoning (\emph{i.e.},
finite structural reasoning) about objects, it is not surprising that
the inductive nominal type $\mu_{F}$ is not supported (nor even an
approximation for it) in most OOP languages. If defined, such a type
or its approximation will contain only objects (in the OO sense, not
the mathematical/category theoretic one) that can be \emph{finitely}
constructed using $F$---which is not of much value (actually, makes
little sense) to most OO software developers\footnote{That is, to developers and designers of industrial-strength and mainstream
OO software, who are content as long as a type that contains all objects
that can be constructed using $F$ (regardless of whether they can
be constructed ``finitely or infinitely'')---namely the free type
$F\left\langle \mathsf{?}\right\rangle $---is defined.}.

\subsection{Structural Induction, Nominal Coinduction, and The Future}

The discussion regarding coinductive nominal types and their approximations,
combined with our observation regarding the coinductiveness of the
subtyping relation (as a set of pairs) in nominal type systems (see
notes of~$\mathsection$\ref{sub:Object-Oriented-Type-Theory}),
make us conclude that fundamental differences exist between structural
type theory (in FP) and nominal type theory (in OOP) regarding, first,
the existence (in FP) versus nonexistence (in OOP) of fixed points
(and pre-/post-fixed points) of type constructors (in FP) and generic
classes (in OOP) and, second, regarding the dominance of (singular)
induction and inductive definitions (in FP) versus the dominance of
(mutual) coinduction and coinductive definitions (in OOP).

In light of the existence of circularity and recursive definitions
in OOP at multiple levels, \emph{i.e.}, at the level of values (\emph{e.g.},
via \code{this}) and at the level of types (\emph{e.g.}, between
class definitions, and noting the circular dependency of parameterized
types on interval types and vice versa), and in light of other work%
{} also discussing coinductive definitions and the use of coinduction
in OOP and in nominal type systems (\emph{e.g.}, see~\cite[p.312]{TAPL}
and~\cite{Ancona2009,Tate2011}), we observe that:
\begin{quote}
``Induction reigns supreme'' in functional programming, while in
object-oriented programming ``coinduction reigns supreme''.
\end{quote}
This observation should not be viewed as a mathematical or academic
statement but rather as an expression of practically-motivated fundamental
features of OOP and FP. The main practical value of functional programming
and the main motivation behind its very existence and behind its continued
usage is to enable precise mathematical reasoning about software.
The main practical value of object-oriented programming and the main
motivation behind its existence and behind its continued usage is
the accurate and intuitive modeling of typically highly-interconnected
real or imaginary parts of our world. We, humans, know well how to
reason, mathematically, about objects built up from few basic ones
(\emph{i.e.}, inductively-defined objects), yet we also need software
to accurately simulate and model parts of our interrelated world---one
in which, to overcome circularity and interrelatedness, we are used
to holding statements about objects as facts as long as these statements
cannot be disproven (\emph{i.e.}, is an inherently coinductive world).

As such, the co-existence of FP (with its mainly inductive type systems)
and OOP (with its mainly coinductive type systems), and our continued
need for both, is a reflection of our reality, \emph{i.e.}, of what
we currently know how to do and of what we currently need to do. It
is our opinion that this current state of affairs is not a perpetual
one, but is one that particularly invites for the further development
of the mathematical methods used to reason about \emph{mutually coinductive
definitions} (and, more generally and more precisely, about mutually-defined
post-fixed points) to reach or exceed the same level of mathematical
maturity of the mathematical methods and tools used to reason about
singular inductive definitions (\emph{i.e.}, about least fixed points).
In summary, the current state of affairs invites PL theorists and
researchers to \emph{make clear and transparent what is }currently\emph{
totally opaque},\footnote{See, for example, the ``totally opaque'' code on p.58 of~\cite{Paulson1996}---a
standard reference on practical functional programming---which nevertheless
has a strong \emph{object-oriented} flavor!} thereby elegantly combining the accurate mathematical-reasoning benefits
of functional programming with the accurate world-modeling benefits
of object-oriented programming.

\bibliographystyle{plain}

\end{document}